\documentclass[11pt,fullpage]{article}

\usepackage{fullpage}
\usepackage{graphicx}
\usepackage{amsmath}
\usepackage{amssymb}
\usepackage{amstext}
\usepackage{amsthm}
\usepackage{epsfig}
\usepackage{placeins}
\usepackage{subcaption}
\usepackage{paralist}
\usepackage{caption}
\usepackage{multirow}
\usepackage{algorithm}
\usepackage{algorithmic}
\usepackage{comment}
\usepackage[obeyspaces,spaces]{url}
\usepackage{balance}
\usepackage{enumitem}
\usepackage{hyperref}
\usepackage{times}

\makeatletter
\newenvironment{sql}%
 {\vskip 5pt\begin{list}{}{%
  \setlength{\topsep}{0pt}\setlength{\partopsep}{0pt}\setlength{\parskip}{0pt}%
  \setlength{\parsep}{0pt}\setlength{\labelwidth}{0pt}%
  \setlength{\rightmargin}{0pt}\setlength{\leftmargin}{0pt}%
  \setlength{\labelsep}{0pt}%
  \obeylines\@vobeyspaces\normalfont\ttfamily%
  \item[]}}
 {\end{list}\vskip5pt\noindent}
\makeatother
\begin{document}

\date{July 2019}

\title{In-Depth Benchmarking of Graph Database Systems with the Linked Data Benchmark Council (LDBC)\\
Social Network Benchmark (SNB)}

\author{Florin Rusu and Zhiyi Huang\\
\{frusu, zhuang29\}@ucmerced.edu\\
University of California Merced
}

\maketitle

%%%%%%%%%%%%%%%%%%%%%%%%%%%%%%%%%%%%%%%%%%%%%%%%%%%%%%%
\begin{abstract}

In this study, we present the first results of a complete implementation of the LDBC SNB benchmark -- interactive short, interactive complex, and business intelligence -- in two native graph database systems---Neo4j and TigerGraph. In addition to thoroughly evaluating the performance of all of the 46 queries in the benchmark on four scale factors -- SF-1, SF-10, SF-100, and SF-1000 -- and three computing architectures -- on premise and in the cloud -- we also measure the bulk loading time and storage size. Our results show that TigerGraph is consistently outperforming Neo4j on the majority of the queries---by two or more orders of magnitude (100X factor) on certain interactive complex and business intelligence queries. The gap increases with the size of the data since only TigerGraph is able to scale to SF-1000---Neo4j finishes only 12 of the 25 business intelligence queries in reasonable time. Nonetheless, Neo4j is generally faster at bulk loading graph data up to SF-100. A key to our study is the active involvement of the vendors in the tuning of their platforms. In order to encourage reproducibility, we make all the code, scripts, and configuration parameters publicly available online.

\end{abstract}

%%%%%%%%%%%%%%%%%%%%%%%%%%%%%%%%%%%%%%%%%%%%%%%%%%%%%%%
\section{INTRODUCTION}\label{sec:intro}

Largely triggered by the proliferation of online social networks over the past decade, there has been an increased demand for processing graph-structured data~\cite{ldbc}. The highly-connected structure of these social networks makes graphs an obvious modeling choice since they provide an intuitive abstraction to represent entities and relationships. As a result, many graph analytics systems and graph databases have been developed both in industry and academia~\cite{ozsu:rdbms-vs-graphs,ldbc-snb}. Graph analytics systems~\cite{graphalg-book}, such as Pregel~\cite{pregel}, Giraph~\cite{giraph}, and GraphLab~\cite{graphlab}, specialize in batch-processing of global graph computations on large computing clusters. On the other hand, graph databases~\cite{graphdb-book}, such as Neo4j~\cite{neo4j}, TigerGraph~\cite{tigergraph}, and Titan/JanusGraph~\cite{janus}, focus on fast querying of relationships between entities and of the graph structure. Graph databases treat relationships as first class citizens and support efficient traversals by native graph storage and indexed access to vertexes and edges. While the traversals are typically expressed by means of imperative APIs due to their complexity, there are also systems that define and implement declarative graph-oriented query languages, such as Neo4j's Cypher~\cite{neo4j-cypher} and TigerGraph's GSQL~\cite{tigergraph-gsql}. However, these graph query languages are not standardized yet~\cite{gcore}, making the use of such systems cumbersome. Nonetheless, this is an important step forward compared to imperative low-level implementation in C++ or Java. Given their superior level of abstraction supported by a declarative query language, graph databases represent the most advanced graph processing systems developed to date.

The plethora and diversity of graph processing engines creates the need for standard benchmarks that help users identify the tools that best suit their applications. Moreover, benchmarks stimulate competition across both academia and industry, which triggers development in the field, e.g., the TPC benchmark suite for relational databases. The Linked Data Benchmark Council (LDBC)~\cite{ldbc} is a joint effort to establish benchmarking practices for evaluating graph data management systems. The main objectives of LDBC are to design benchmark specifications and procedures, and publish benchmarking results~\cite{ldbc-paper}. The LDBC Social Network Benchmark (SNB)~\cite{ldbc-snb} is the first result of this effort. It models a social network graph and introduces two different workloads on this common graph. The Interactive Workload~\cite{snb-interactive} specifies a set of read-only traversals that touch a small portion of the graph and is further divided into interactive short (IS) and interactive complex (IC) queries. The Business Intelligence (BI) Workload~\cite{snb-bi} explores large portions of the graph in search of occurrences of patterns that combine both structural and attribute predicates of varying complexity. This is different from graph analytics workloads~\cite{graphalytics} in that the identified patterns are typically grouped, aggregated, and sorted to summarize the results. Given the common underlying graph structure and the extensiveness of graph algorithms embedded in the query workloads, LDBC SNB is the most complete benchmark for graph databases to date.

%%%%%%%%%%%%%%%%%%%%%%%%%
\paragraph*{Contributions.}
We present the first exhaustive results of a complete implementation of the LDBC SNB benchmark in two native graph database systems with declarative language support---Neo4j and TigerGraph. We have implemented all of the 46 queries in the benchmark both in Neo4j's Cypher and TigerGraph's GSQL query languages, and optimized them with direct input from the system developers. These query statements have been used for the successful cross-validation of the two query languages and can be taken as reference for future implementations. To this end, we make all the code, scripts, and configuration parameters publicly available online in order to encourage reproducibility. We evaluate query performance over four scale factors -- ranging from SF-1 (1 GB) to SF-1000 (1 TB) -- on three computing architectures -- on premise and in the cloud -- exhibiting large variety in terms of number of CPUs and memory capacity. Additionally, we also measure the bulk loading time and storage size. Our results show that TigerGraph is consistently outperforming Neo4j on the majority of the queries---by two or more orders of magnitude on certain interactive complex and business intelligence queries. The gap increases with the size of the data. Moreover, only TigerGraph is able to scale to SF-1000 since Neo4j finishes only 12 of the 25 business intelligence queries in reasonable time. However, Neo4j is generally faster at bulk loading graph data up to SF-100, even though indexing has to be performed as an explicit additional process.

%%%%%%%%%%%%%%%%%%%%%%%%%%%%%%%%%%%%%%%%%%%%%%%%%%%%%%%
\section{LDBC SNB BENCHMARK}\label{sec:ldbc-snb}

In this section, we provide a short introduction of the LDBC SNB benchmark. We present the schema, the data generation process, and the query workload. A thorough presentation of the benchmark is available in the original publications~\cite{snb-interactive,snb-bi} and the official specification~\cite{ldbc-snb,ldbc-snb-doc,ldbc-snb-datagen}.

%%%%%%%%%%%%%%%%%%%%%%%%%
\begin{figure}[htbp]
\begin{center}
 \includegraphics[width=\textwidth]{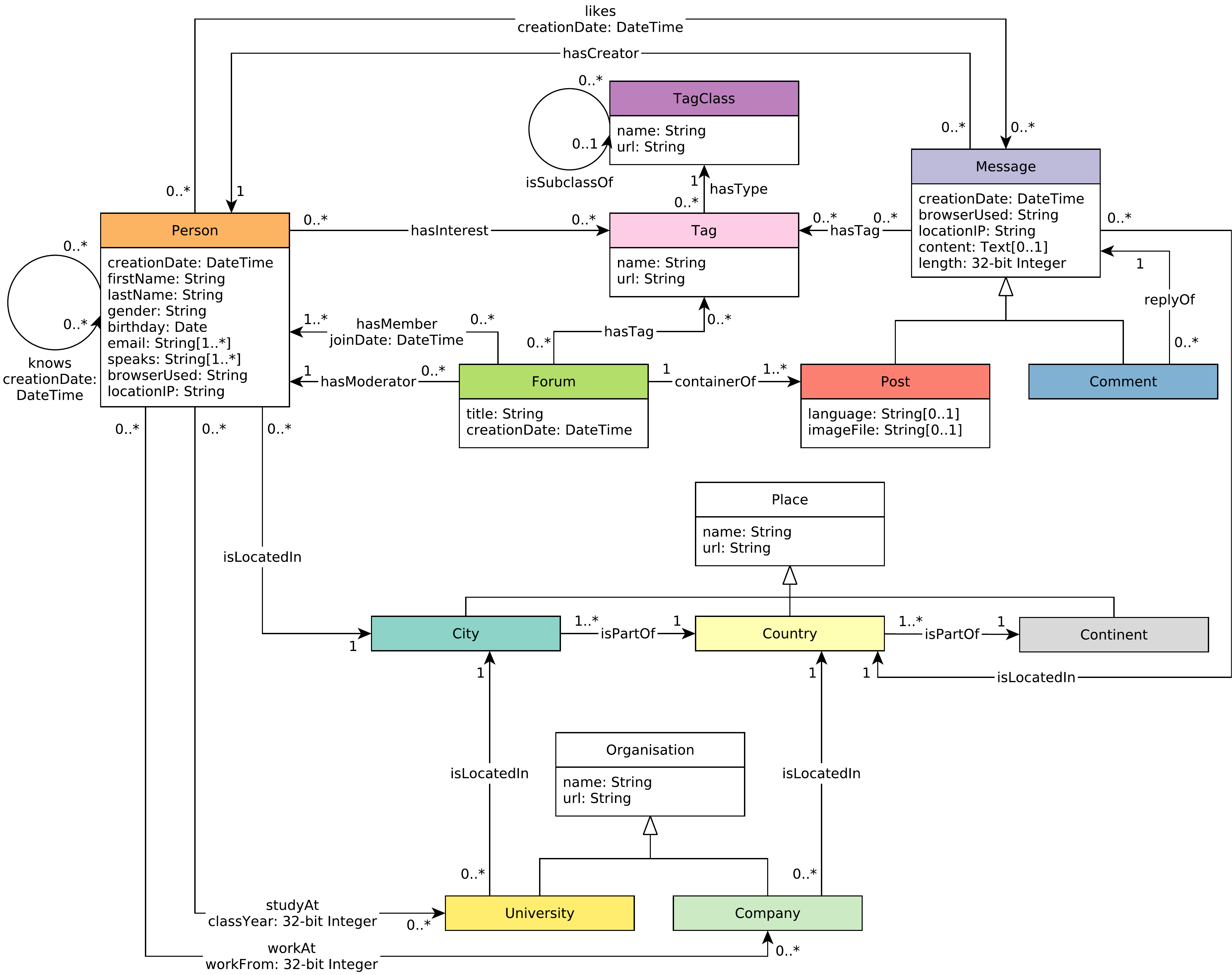}
\end{center}
\caption{The LDBC SNB data schema (reproduced exactly from~\cite{ldbc-snb-doc}).}
\label{fig:snb-schema}
\end{figure}
%%%%%%%%%%%%%%%%%%%%%%%%%

%%%%%%%%%%%%%%%%%%%%%%%%%
\paragraph*{Schema.}
The schema of the LDBC SNB benchmark is depicted as a UML diagram in Figure~\ref{fig:snb-schema}. The schema represents a realistic social network, including people and their activities, over a period of time. It defines the structure of the data in terms of entities and their relationships. The main entity is Person. Each person has a series of attributes, such as name, gender, and birthday, and a series of relationships with other entities in the schema, e.g., a person graduated from a university in a certain year. A person's activity is represented in the form of friendship relationships with other persons and content sharing, such as messages, replies to messages, and likes of messages. Persons form groups, i.e., forums, to talk about specific topics---represented as tags. The dataset generated from the schema forms a graph that is a fully connected component of persons over their friendship relationships. Each person has a few forums under which the messages form large discussion trees. The messages are connected to persons by authorship and likes. Organization and Place information are dimension-like and do not scale with the number of persons. Time is an implicit dimension represented as DateTime attributes connected to entities and relationships. While the structure of the LDBC SNB schema is a graph, the benchmark does not enforce any particular physical representation. This allows for storage both as tables in a relational databases, as well as graphs in a graph database.

%%%%%%%%%%%%%%%%%%%%%%%%%
\paragraph*{Data.}
The LDBC SNB data generator instantiates synthetic datasets of different scale with distributions and correlations similar to those expected in a real social network. This is realized by integrating correlations in attribute values, activity over time, and the graph structure~\cite{snb-interactive,ldbc-snb-datagen}. The attribute values are extracted from the DBpedia dictionary. They are correlated among themselves and also influence the connection patterns in the social graph. For example, the location where a person lives influences their name, university, company, and spoken languages, their interests, i.e., forums and tags, which, in turn, influence the topics of his/her posts, which also influence the text of the messages. The volume of a person's activity, i.e., number of messages, is driven by real world events. Whenever an important event occurs, the amount of people and messages talking about that topic spikes---especially from those persons interested in that topic. The graph structure is dependent on the attribute values of the connected entity instances. For example, persons that are interested in a topic and have studied in the same university during the same year, have a larger probability to be friends. Similar to influencers and communities, the number of friends is skewed across persons. Moreover, the correlations in the friends graph also propagate to messages and comments. The scale of the benchmark is driven by the number of persons in the social network which directly impacts all the other cardinalities. However, since the generation process is so complicated, there is no clear correspondence between datasets of different scale factors. Specifically, it is not the case that a larger dataset includes a smaller one. This poses difficulty in selecting the query parameters because result cardinality is not increasing linearly with the scale factor. It is possible to get results with lower cardinality -- or even empty results -- when executing the same query on a larger scale factor.

%%%%%%%%%%%%%%%%%%%%%%%%%
\paragraph*{Queries.}
Three query workloads are defined over the common SNB schema and corresponding data. They are interactive short (IS), interactive complex (IC), and business intelligence (BI). Each of the workloads consists of a number of query templates -- 7 for IS, 14 for IC, and 25 for BI -- that test different characteristics of the system implementing the benchmark. In order to be as close as possible to real world scenarios, the choice of the queries is driven by choke points -- aspects of query execution or optimization which are known to be problematic -- extracted from real systems. Examples include estimating cardinality in graph traversals with data skew and correlations; handling scattered index access patterns; sub-query, intra-query, and inter-query result reuse; top-k push down; late projection; sparse foreign key joins; dimensional clustering; etc. The complete list is available in~\cite{snb-bi}. Each query template comes together with a set of predefined substitution values for the template parameters, i.e., parameter bindings. Given the skewed structure of the graph, the choice of the parameter bindings requires special attention because it can result in execution times that exhibit high variance. A parameter curation process~\cite{snb-interactive} that selects the substitution values during data generation is designed to ensure stable execution times across all parameter bindings. Due to the structural graph changes, a different series of bindings is generated for each scale factor.

%%%%%%%%%%%%%%%%%%%%%%%%%
\paragraph*{Interactive Short (IS) Workload.}
The queries in the IS workload are relatively simple path traversals that access at most 2-hop vertexes from the origin---given as a parameter binding. They originate either from a person or a message and access the friendship neighborhood and the associated messages. Assuming that the origin can be retrieved with an index lookup, the amount of data that has to be accessed is considerably smaller than the dataset size. Moreover, the execution time is not directly impacted by the scale factor since the same amount of data is accessed.

We consider query IS\_3 as an example: \textit{given a person identified by the id, find all their friends and the date at which they became friends; return the friends from the most to the least recent}. This query requires a lookup by the person id to find the friends, followed by another lookup for each friend to find their information. Finally, sorting is performed on the list of friends which should have relatively small cardinality. The amount of data that is accessed is proportional to the number of friends of the input person---retrieved on the 1-hop friendship path from the origin person. The other IS queries have from 0 to 2 hops. Their specification can be found in~\cite{ldbc-snb-doc}.

%%%%%%%%%%%%%%%%%%%%%%%%%
\paragraph*{Interactive Complex (IC) Workload.}
The queries in the IC workload go beyond 2-hop paths and compute simple aggregates---rather than returning only tuples. Additionally, two of the queries have to calculate the shortest path between two vertexes given as parameters. In order to support such queries, recursion has to be an integral part of the execution engine because the depth of the traversal depends on the parameter binding. This eliminates preprocessing and materialization as a viable solution. Although the origin of the traversal is still fixed, the amount of data that a query has to examine is larger than for IS queries. Moreover, aggregate computation requires state handling at global or group level. These generally result in an increase of the execution time with the scale factor.

We consider query IC\_9 as a representative example for this workload: \textit{given a person identified by the id, find the most recent 20 messages posted by their friends or friends of friends before a given date}. This query looks for paths of length two or three, starting from a given person, moving to its friends and friends of friends, and ending at their posted messages. While friends are typically materialized, finding friends of friends requires an additional graph traversal. The result of this traversal is merged with the direct friends and used for the traversal of the messages. If friendships with a higher degree are allowed, more traversals are required. The date parameter can be used to prune the number of considered messages. The order in which to perform the traversals over friends and messages makes a significant difference in execution time. This is also the case for the other IC queries~\cite{snb-interactive}.

%%%%%%%%%%%%%%%%%%%%%%%%%
\paragraph*{Business Intelligence (BI) Workload.}
The queries in the BI workload access a much larger part of the graph. This is realized by replacing the origin in the interactive queries with more general selections on message creation date or the location of a person. As a result, traversals are not originating from a single source, but rather from multiple points in the graph. The aggregates that have to be computed are also more complicated. They involve complex grouping criteria over multiple attributes -- some of them synthesized -- and non-trivial functions. Top-k ranking and sorting are applied over these complex aggregates. In addition to the single source -- single destination shortest paths, more general weighted paths and fixed size cliques, e.g., triangles, are part of the BI workload. The efficient execution of the BI queries requires extensive optimizations across all the layers of the system---not only graph traversal.

We take BI\_5 as a representative query for this workload: \textit{find the 100 most popular forums in a given country; for each person in these forums, count the number of posts they made across all of the popular forums}. This query is a good combination of graph traversal and complex top-k aggregation. The optimal graph traversal requires finding the proper direction between forums and persons. Top-k is applied as a condition to a count aggregate and the result is further used in another group-by aggregate. This requires advanced support both at the query language, as well as in the execution engine. The other queries in the BI workload follow the same pattern and level of complexity~\cite{snb-bi,ldbc-snb-doc}.

%%%%%%%%%%%%%%%%%%%%%%%%%%%%%%%%%%%%%%%%%%%%%%%%%%%%%%%
\section{DECLARATIVE GRAPH DATABASE SYSTEMS}\label{sec:dec-graph-db}

In this section, we introduce the two graph database systems -- Neo4j~\cite{neo4j,graphdb-book} and TigerGraph~\cite{tigergraph,tigergraph-system-arxiv} -- considered in this work, focusing on their query languages -- Cypher~\cite{neo4j-cypher,graphalg-book} and GSQL~\cite{tigergraph-gsql,tigergraph-language-arxiv}, respectively -- and not on their architecture or execution engine. Our goal is to provide complete implementations of the LDBC benchmark as declarative graph queries and analyze their characteristics. The reason we choose Neo4j and TigerGraph is because they are the only native graph databases freely available that provide a SQL-like declarative query language. We do not consider systems such as Oracle PGX~\cite{oracle-pgql} and Amazon Neptune~\cite{neptune} which are not free or can run only in the cloud. We also do not consider systems that support the Gremlin~\cite{gremlin} functional/procedural query language, e.g., JanusGraph~\cite{janus}, or implementations of graph operations as SQL/SPARQL statements in relational and RDF databases.

%%%%%%%%%%%%%%%%%%%%%%%%%
\paragraph*{Cypher.}
Neo4j's graph query language is based on the \textit{labeled property graph data model}~\cite{graphdb-book,gcore}---which is the most popular model for representing graphs. A labeled property graph is a directed graph with labels on both vertexes and edges, as well as $<$property,value$>$ pairs associated with both. Typically, there are multiple properties associated with a vertex/edge, while there is a single label. In relational terms, the label corresponds to the table name, while the properties correspond to the attributes of the table. However, each vertex/edge can have its own independent label and properties, i.e., they are not tuples from a predefined table. Thus, the labeled property graph model is schema-less. On one hand, this provides extreme flexibility, on the other, it enlarges the storage and makes evaluation less efficient. Cypher queries are clauses that specify paths in the graph. They identify vertexes by their label and restrict the -- possibly large -- set of matches by their properties. The edges on the path are also selected based on their label. Following a declarative style, Cypher defines a limited number of keywords---many of them borrowed from SQL. Clauses are composed based on the Datalog syntax. 

To make things concrete, we provide the Cypher statements for LDBC queries IS\_3 and IC\_9---the statements for all the LDBC queries are available online~\cite{ldbc-snb-code,zhiyi-code}. IS\_3 is a basic \texttt{MATCH-RETURN} Cypher clause that starts from a \texttt{Person} vertex with a given \texttt{id} and matches all the paths that have a single \texttt{Knows} edge \texttt{r} leading to another \texttt{Person} vertex referenced by variable \texttt{friend}. The query returns properties of \texttt{friend} and \texttt{r} in sorted order. The output can be seen as a relational table with a fixed schema. Query IC\_9 specifies a more complicated path. First, it includes friends-of-friends, i.e., \texttt{Person} vertexes connected by two \texttt{Knows} edges. Second, the directed edges \texttt{Has\_Creator} to a \texttt{Message} posted by the \texttt{friend} are matched. Only those \texttt{Message} vertexes posted before \texttt{maxDate} are considered.

%%%%%%%%%%%%%%%%%%%%%%%%%
\begin{sql}
MATCH (:Person {id:\$personId})-[r:Knows]-(friend:Person)
RETURN
	\hspace*{0.5cm} friend.id AS personId,
	\hspace*{0.5cm} friend.firstName AS firstName,
	\hspace*{0.5cm} friend.lastName AS lastName,
	\hspace*{0.5cm} r.creationDate AS friendshipCreationDate
ORDER BY friendshipCreationDate DESC, personId ASC
\end{sql}\label{cypher:is-3}
%%%%%%%%%%%%%%%%%%%%%%%%%

%%%%%%%%%%%%%%%%%%%%%%%%%
\begin{sql}
MATCH (:Person {id:\$personId})-[:Knows*1..2]-(friend:Person)
	\hspace*{1.2cm} <-[:Has\_Creator]-(message:Message)
WHERE message.creationDate < \$maxDate
RETURN DISTINCT
	\hspace*{0.5cm} friend.id AS personId,
	\hspace*{0.5cm} friend.firstName AS personFirstName,
	\hspace*{0.5cm} friend.lastName AS personLastName,
	\hspace*{0.5cm} message.id AS messageId,
	\hspace*{0.5cm} CASE exists(message.content)
		\hspace*{1cm} WHEN true THEN message.content
		\hspace*{1cm} ELSE message.imageFile
	\hspace*{0.5cm} END AS messageContent,
	\hspace*{0.5cm} message.creationDate AS messageCreationDate
ORDER BY messageCreationDate DESC, messageId ASC
LIMIT 20
\end{sql}\label{cypher:ic-9}
%%%%%%%%%%%%%%%%%%%%%%%%%

These queries show the close relationship between Cypher and SQL. In fact, Cypher inherits the expressiveness of SQL, i.e., it is SQL-complete. This is realized by the composition of the labeled property graph model and table functions. As a result, Cypher has very limited control flow support and query composition through subqueries is rather difficult. These limit the graph computations that can be expressed in Cypher---especially, recursive and iterative algorithms. Nonetheless, due to its resemblance to SQL, Cypher represents a relatively easy transition to Neo4j.

%%%%%%%%%%%%%%%%%%%%%%%%%
\paragraph*{GSQL.}
GSQL is the TigerGraph query language~\cite{tigergraph-system-arxiv}. As the name suggests, GSQL~\cite{tigergraph-gsql} is a direct extension of SQL to graph databases. It imposes a strict schema declaration before querying. The schema implements the labeled property graph data model and consists of four types---vertex, edge, graph, and label~\cite{tigergraph-language-arxiv}. The vertex type corresponds to a SQL table. It has a name and attributes. The edge type is defined between two vertex types. It can be undirected or directed. In the case of a directed edge, an optional reverse edge type can be defined. The graph type defines the vertex and edge types that create the graph. The label type is included only for compatibility with the labeled graph data model. Since everything is specified from the beginning, TigerGraph can employ the optimal storage format and query execution strategy. Queries in GSQL are not single SQL \texttt{SELECT} statements, but rather stored procedures consisting of multiple \texttt{SELECT} clauses and imperative instructions such as branches and loops. Essentially, a GSQL query is a SQL stored procedure. The motivation for this approach is the increased complexity of certain graph computations. Similar to \texttt{MATCH} in Cypher, the \texttt{SELECT} statement in GSQL matches a path in the graph starting from a vertex and following edges. The path is specified in the \texttt{FROM} clause. GSQL introduces the concept of accumulator \texttt{ACCUM} associated with a path. The data found along a path can be collected and aggregated into  accumulators according to distinct grouping criteria. This is done in parallel, with one thread for each match in the \texttt{FROM} clause. The aggregated results can be distributed across vertexes in order to support multi-pass and iterative computations. 

We include the GSQL statements for LDBC queries IS\_3 and IC\_9 as a comparison reference to the corresponding Cypher queries. The GSQL implementation for all the other LDBC queries is available online~\cite{tigergraph-ldbc-snb-queries}. The \texttt{SELECT} statement in IS\_3 is very similar to the \texttt{MATCH} statement in Cypher. The main difference is the accumulator \texttt{creationDate} attached to the \texttt{Person} vertex returned by \texttt{SELECT}. This is necessary because \texttt{SELECT} returns a well-defined type in GSQL. IC\_9 showcases multiple GSQL features, including several types of accumulators and loops. The friends and friends-of-friends of the parameter \texttt{Person} are computed within a procedural loop that traverses two \texttt{Person\_Knows\_Person} edges from the origin. They are all stored in the set accumulator \texttt{friendAll}. The \texttt{Messages} posted by the friends are stored in a heap accumulator of fixed size with a comparison function declared at definition. This allows only the relevant \texttt{Messages} to be considered. Compared to the Cypher code, GSQL is not as concise because it includes imperative instructions. However, these follow the well-established stored procedure SQL paradigm which embeds all the logic as a compiled object inside the database. The application has only to invoke the procedure through a function call. Although the example queries do not show the complete GSQL expressiveness, there are many graph computations that cannot be expressed directly in Cypher, however, they can be written as GSQL queries---while Cypher is SQL-complete, GSQL is Turing-complete.

%%%%%%%%%%%%%%%%%%%%%%%%%
\begin{sql}\label{gsql:is-3}
CREATE QUERY IS\_3(VERTEX<Person> personId) FOR GRAPH ldbc\_snb \{
  SumAccum<INT> @creationDate;

  vPerson = \{\$personId\};
  vFriend =	SELECT t
  					\hspace*{1.84cm} FROM vPerson:s-(Person\_Knows\_Person:e)->Person:t
				    \hspace*{1.64cm} ACCUM t.@creationDate+=e.creationDate
				    \hspace*{1.64cm} ORDER BY t.@creationDate DESC, t.id ASC;

  PRINT vFriend [
      vFriend.id AS personId,
      vFriend.firstName AS firstName,
      vFriend.lastName AS lastName,
      vFriend.@creationDate AS friendshipCreationDate
	\hspace*{0.2cm} ];
\}
\end{sql}
%%%%%%%%%%%%%%%%%%%%%%%%%

%%%%%%%%%%%%%%%%%%%%%%%%%
\begin{sql}\label{gsql:ic-9}
CREATE QUERY IC\_9(VERTEX<Person> personId, DATETIME maxDate)
FOR GRAPH ldbc\_snb \{
  TYPEDEF tuple < INT personId,
  				  STRING personFirstName,
  				  STRING personLastName,
  				  INT messageId,
  				  STRING messageContent,
  				  DATETIME messageCreationDate
  				> msgInfo;

  OrAccum @visited;
  SetAccum<VERTEX<Person>> @@friendAll;
  HeapAccum<msgInfo>(
	\hspace*{0.7cm} 20, messageCreationDate DESC, messageId ASC
  ) @@msgInfoTop;

  vPerson = \{\$personId\};

  INT i = 0;
  WHILE i < 2 DO
    vPerson = SELECT t
		\hspace*{3cm} FROM vPerson:s-(Person\_Knows\_Person:e)->Person:t
      	\hspace*{1.4cm} WHERE t.@visited==False
      	\hspace*{1.4cm} ACCUM s.@visited+=True, t.@visited+=True, @@friendAll+=t;

    i = i + 1;
  END;

  vFriend = \{@@friendAll\};
  vMessage =
  	SELECT t
	\hspace*{0.5cm} FROM vFriend:s-(Comment\_Has\_Creator\_Person\_REVERSE:e)->Comment:t
    \hspace*{-0.4cm} WHERE t.creationDate < \$maxDate
    \hspace*{-0.4cm} ACCUM @@msgInfoTop += msgInfo(s.id, s.firstName, s.lastName,
    \hspace*{2cm} t.id, t.content, t.creationDate);

  PRINT @@msgInfoTop;
\}
\end{sql}
%%%%%%%%%%%%%%%%%%%%%%%%%

%%%%%%%%%%%%%%%%%%%%%%%%%%%%%%%%%%%%%%%%%%%%%%%%%%%%%%%
\section{BENCHMARK EXPERIMENTS}\label{sec:benchmark}

In this section, we present the results for executing the complete LDBC SNB benchmark in Neo4j and TigerGraph. While results for a subset of the workloads have been presented before -- IS and IC in~\cite{snb-interactive}, BI in~\cite{snb-bi} -- this is the first work that considers all the workloads in a single place. Moreover, we are the first to present results for scale factor SF-1000 which corresponds to 1 TB of data. Our main focus is to report query execution times for the two systems. Additionally, we also evaluate data loading performance in terms of loading time and storage size. Before we present the results, we first introduce the experimental setup.

%%%%%%%%%%%%%%%%%%%%%%%%%%%%%%%%%%
\subsection{Setup}\label{sec:experiments:setup}

%%%%%%%%%%%%%%%%%%%%%%%%%%%%%%%%%%
\paragraph*{Implementation.}
The two graph databases used in the experiments are \textit{Neo4j 3.5.0 Community Edition} and \textit{TigerGraph 2.3.1 Developer Edition}. These versions are available for free and may not include all the optimizations provided in commercially supported versions. While both systems can run in distributed mode, we configure them optimally for single-node execution, i.e., we allow full memory and thread utilization. The Cypher queries are implemented in Python and passed for execution to the Neo4j server over a standard ODBC/JDBC connection. The exact statements from the LDBC repository~\cite{ldbc-snb-code} are used. The results and timing measurements are returned to the Python application for logging/display. The process to execute queries in TigerGraph follows the stored procedure workflow from relational databases. First, the query has to be registered and compiled in the server. This creates a database object registered in the catalog associated with the executable code corresponding to the query. In the second stage, the query/stored procedure is invoked through a function call by a Python application similar to the Neo4j driver. Due to its relative novelty, the GSQL query statements~\cite{tigergraph-ldbc-snb-queries} have been optimized with direct input from TigerGraph engineers to whom we thank for their help. We provided the same opportunity for optimizations to Neo4j staff, however, they declined to give us any input for more than two months at the time of this publication. The query results of the two systems have been used for the successful cross-validation of the Cypher and GSQL query languages and can be taken as reference for future implementations of the LDBC benchmark.

%%%%%%%%%%%%%%%%%%%%%%%%%%%%%%%%%%
\paragraph*{Systems.}
We use three different machines to perform the experiments. Their properties are given in Table~\ref{tbl:specs}. The smallest machine is our server at UC Merced which has been dedicated exclusively to run the two graph databases. The other two machines are Amazon AWS instances. While AWS r4.8xlarge is virtual and can be shared by multiple users, AWS x1e.16xlarge is a physical instance that requires exclusive reservation. Ubuntu 18.04.2 LTS is the operating system on all the machines. Based on the available memory capacity, we perform the experiments for a given scale factor on the smallest machine with sufficient memory to support the required data size. Thus, SF-1 (1 GB) and SF-10 (10 GB) are executed on the UC Merced server, SF-100 (100 GB) is executed on AWS r4.8xlarge, and SF-1000 (1 TB) is executed on AWS x1e.16xlarge, respectively. In addition to the difference in memory size, the three machines exhibit a large variation in the number of CPU cores/threads. This can result in significantly different degrees of parallelism, especially for queries that access similar amounts of data independently of the scale factor, e.g., the IS queries. Overall, the combination of data size and hardware characteristics provides an extensive picture of the performance of the two graph databases on the LDBC benchmark. We are not aware of any other study that takes such an exhaustive approach.

%%%%%%%%%%%%%%%%%%%%%%%%%%%%%%%%%%
\begin{table}[htbp]
  \begin{center}

  \resizebox{\textwidth}{!}{
		\begin{tabular}{l|l|r|r|l|l|l}
		
		\hline
		
		\textbf{Scale factor} & \textbf{Machine} & \textbf{(virtual) CPU cores} & \textbf{RAM} & \textbf{OS} & \textbf{Java} & \textbf{Python} \\
		
		\hline\hline
		SF-1 \& SF-10	& UC Merced	& 16	& 28 GB	& Ubuntu 18.04.2 LTS	& build 1.8.0\_191	& 2.7.15 \\
		SF-100	& AWS r4.8xlarge	& 32	& 240 GB	& Ubuntu 18.04.2 LTS	& build 1.8.0\_191	& 2.7.15 \\
		SF-1000	& AWS x1e.16xlarge	& 64	& 2 TB	& Ubuntu 18.04.2 LTS	& build 1.8.0\_201-b09 &	3.6.5 \\
		
		\hline
	
    \end{tabular}
  }
  
  \end{center}
\caption{Specification of the hardware and system software used for benchmarking.}\label{tbl:specs}
\end{table}
%%%%%%%%%%%%%%%%%%%%%%%%%%%%%%%%%%

%%%%%%%%%%%%%%%%%%%%%%%%%%%%%%%%%%
\paragraph*{Methodology.}
By default, we perform all the queries 10 times and report as result -- depicted in the charts and tables -- the median of the last 9 runs. We use the median instead of the mean because it is more stable to rare variations in runtime. The first run is ignored because it typically takes much longer than the subsequent ones. This is due to the cold cache startup. In order to limit the amount of time dedicated to a specific query, we impose a timeout of 18,000 seconds (5 hours). When the timeout expires, the query is terminated. The data loading experiments are performed only once since they take much longer and have a more stable runtime. The data size reported in the storage experiments is measured from the space occupied on the file system with the \texttt{ls} and \texttt{du} commands.

%%%%%%%%%%%%%%%%%%%%%%%%%%%%%%%%%%
\paragraph*{Datasets.}
The characteristics of the LDBC SNB graphs used in the experiments are given in Table~\ref{tbl:datasets}. We can observe the almost linear increase in the number of vertexes and edges with the scale factor. The relationship is not exactly linear because the data generation process is taking the sparsity of the graph in consideration---the degree of the nodes is kept relatively constant as the scale factor increases. Nonetheless, at SF-1000, the graph has almost 2.7 billion vertexes and 18 billion edges, which require 900 GB storage in raw format. By all accounts, this is an extremely large graph even for the largest existing social networks.

%%%%%%%%%%%%%%%%%%%%%%%%%%%%%%%%%%
\begin{table}[htbp]
  \begin{center}

		\begin{tabular}{l|rr|r}
		
		\hline
		
		\textbf{Scale factor} & \textbf{Vertexes (millions)} & \textbf{Edges (millions)} & \textbf{Raw size (GB)} \\
		
		\hline\hline
		SF-1 	& 3.18 		& 17.26 	& 0.813 \\
		SF-10	& 30.00 	& 176.62 	& 8.400 \\
		SF-100	& 282.64 	& 1,780.00 	& 87.300 \\
		SF-1000	& 2,690.00 	& 17,790.00	& 900.000 \\
		
		\hline
	
    \end{tabular}
  
  \end{center}
\caption{LDBC SNB graph characteristics.}\label{tbl:datasets}
\end{table}
%%%%%%%%%%%%%%%%%%%%%%%%%%%%%%%%%%

%%%%%%%%%%%%%%%%%%%%%%%%%%%%%%%%%%
\subsection{Results}\label{sec:experiments:results}

We group the results into loading and querying. For loading, we measure the storage size required by the graphs in the two systems and the time to load the data before it is available for querying, i.e., time-to-query. For querying, we report the runtime to perform all the 46 SNB queries across the four scale factors in Neo4j and TigerGraph---for a total of 368 configurations.

%%%%%%%%%%%%%%%%%%%%%%%%%%%%%%%%%%
\paragraph*{Loaded data size.}
The raw graphs generated by the LDBC SNB data generator are bulk loaded into the two graph databases using the supported constructs in their respective query language. Neo4j provides import APIs that build the labeled property graph from different formats. We use the import API from separator delimited text files. This API loads vertexes and edges having the same label with a single command, e.g., all the \texttt{Person} vertexes are loaded at once, and it loads all the properties of each vertex/edge instance. In order to identify vertexes having a specified property efficiently, indexes have to be built on the loaded data. This is a separate post-loading process. Based on the SNB workloads, we create the following indexes in Neo4j: \texttt{Person(id)}, \texttt{Message(id)}, \texttt{Post(id)}, \texttt{Comment(id)}, \texttt{Forum(id)}, \texttt{Organisation(id)}, \texttt{Place(name)}, \texttt{Tag(name)}, and \texttt{TagClass(name)}. TigerGraph acknowledges the complexity of loading graph data and provides a specialized Data Loading Language (DLL) in GSQL. Thus, loading in TigerGraph is performed with declarative GSQL statements derived from the graph definition DDL. These statements create parallel jobs to extract the vertex/edge properties and ingest them into the internal TigerGraph storage format. There is no pre-processing, e.g., extracting unique vertex ids, or post-processing, e.g., explicit index building. Since the entire process is performed offline, this is called offline loading in TigerGraph. There are two stages in offline loading---load and build. The load stage parses and prepares the binary data, while build merges and packages the binary data into the actual storage format.

%%%%%%%%%%%%%%%%%%%%%%%%%
\begin{figure}[htbp]
\begin{center}
 \includegraphics[width=\textwidth]{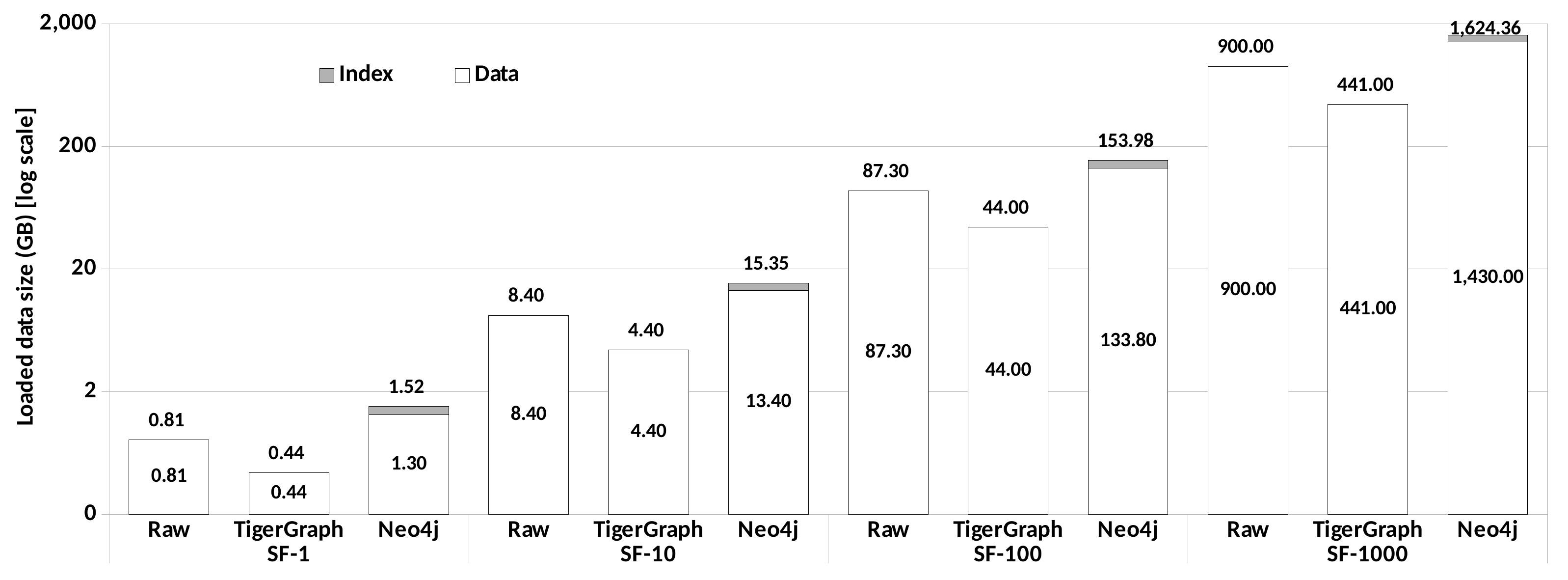}
\end{center}
\caption{Loading data size split into actual data size and indexes size. Raw corresponds to the size of the data generated by the LDBC benchmark generator. TigerGraph does not create explicit indexes. The numbers inside the bars represent the size of the actual data, while the numbers on top of the bars correspond to the total data size. The difference between the two represents the size of the indexes.}
\label{fig:load-data}
\end{figure}
%%%%%%%%%%%%%%%%%%%%%%%%%

Figure~\ref{fig:load-data} depicts the size of the loaded data -- as well as the original raw data -- for all the considered scale factors. We can observe that TigerGraph approximately halves the size of the raw data, while Neo4j doubles it. The compression achieved by TigerGraph is due to the fixed format imposed by the graph schema which eliminates the need to store the property name for every instance. In Neo4j, this results in more than 50\% increase in size over the raw data. The 9 indexes add the remaining 40\% for the approximately doubling in size. Thus, it is clear that imposing a graph schema has a positive impact on storage. The size of the data in TigerGraph is 3X -- if we include the indexes 4X -- smaller than in Neo4j.

\begin{figure}[htbp]
\begin{center}
 \includegraphics[width=\textwidth]{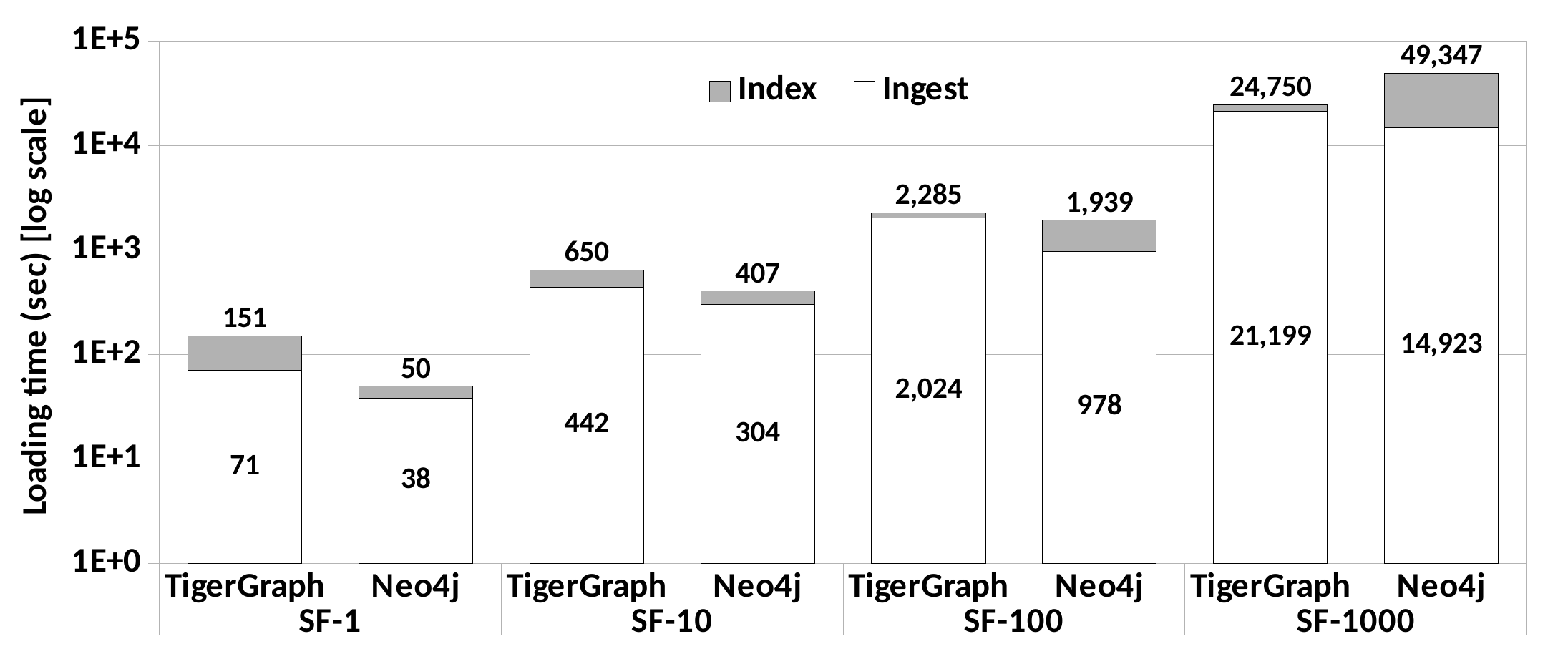}
\end{center}
\caption{Loading time split into ingestion time and indexing time. The numbers inside the bars represent the ingestion time, while the numbers on top of the bars correspond to the total loading time. The difference between the two represents the indexing time---build stage time in TigerGraph.}
\label{fig:load-time}
\end{figure}

%%%%%%%%%%%%%%%%%%%%%%%%%%%%%%%%%%
\paragraph*{Loading time.}
Figure~\ref{fig:load-time} depicts the loading time split into the time to ingest the data and the time to create the indexes. Measuring this time in Neo4j requires timing two separate processes. Even so, the total loading time in Neo4j is smaller than in TigerGraph---except for SF-1000. The difference between the two systems decreases with the increase in scale factor. This is entirely due to indexing efficiency. Index building is not scalable in Neo4j, increasing exponentially with the scale factor---from 12 (SF-1) to 103 (SF-10) to 961 seconds (SF-100). For SF-1000, indexing takes 34,424 seconds, which is more than twice the ingestion time. Since TigerGraph does not have indexing, we consider the time taken by the build stage of offline loading as the equivalent of indexing in Neo4j. The build stage is considerably more scalable than indexing, taking only 3,551 seconds for SF-1000. This huge difference accounts for the smaller loading time in TigerGraph, even though the ingestion time in Neo4j is still only two thirds of the time in TigerGraph. While indexing time in Neo4j could be reduced by building fewer indexes, we will see that -- even with such a large number of indexes -- querying SF-1000 data is very inefficient. With fewer indexes, it is likely that almost no queries would finish in acceptable time.

%%%%%%%%%%%%%%%%%%%%%%%%%
%%%%%%%%%%%%%%%%%%%%%%%%%

%%%%%%%%%%%%%%%%%%%%%%%%%
\begin{figure}[htbp]
\centering

\begin{subfigure}[b]{.76\textwidth}
 \includegraphics[width=\linewidth]{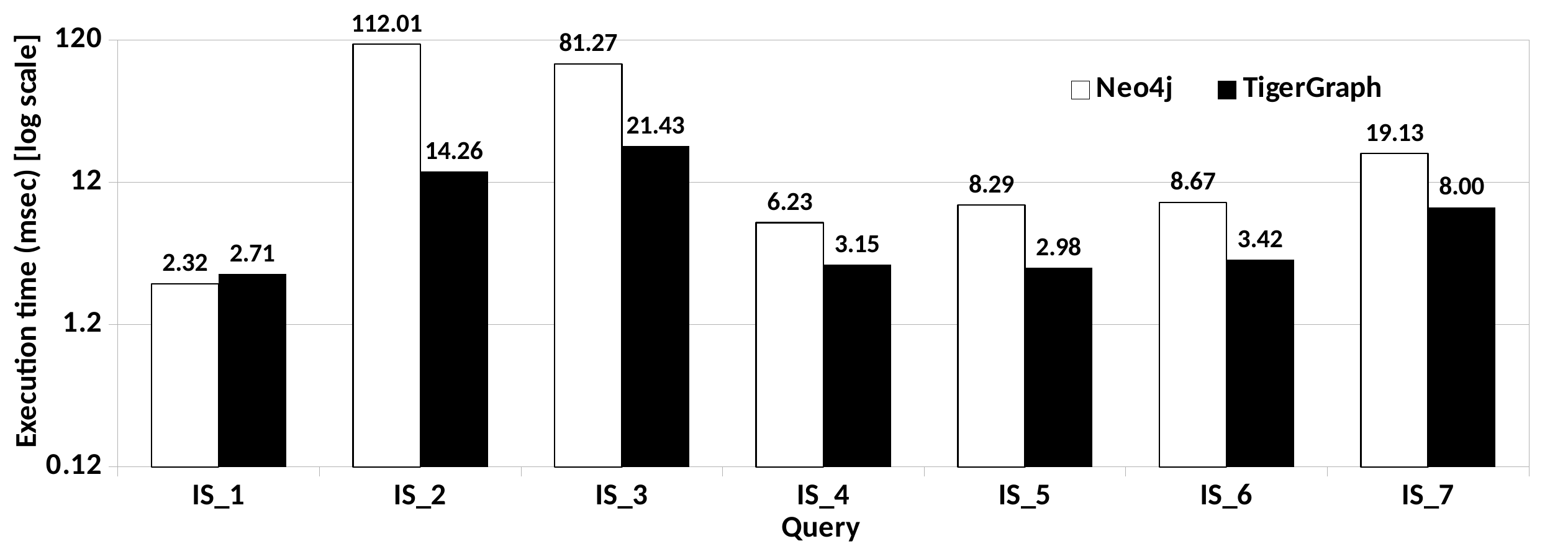}
 \caption{}
 \label{fig:sf-1-is}
\end{subfigure}

\begin{subfigure}[b]{.76\textwidth}
 \includegraphics[width=\linewidth]{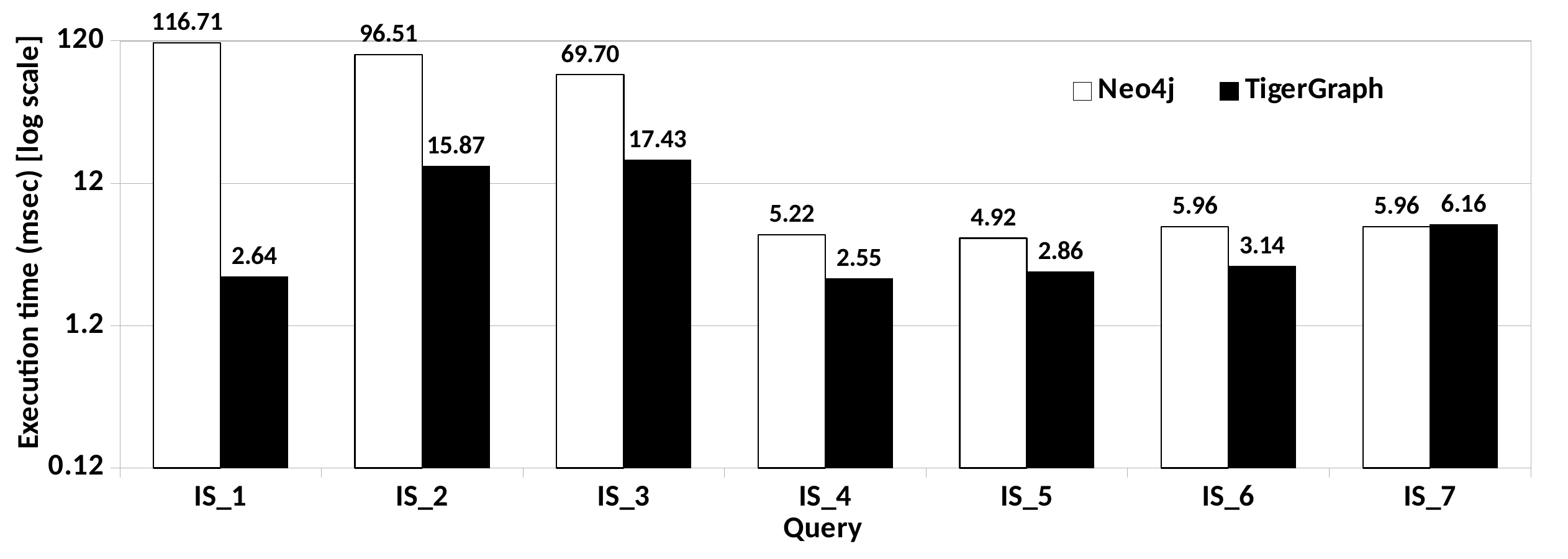}
 \caption{}
 \label{fig:sf-10-is}
\end{subfigure}

\begin{subfigure}[b]{.76\textwidth}
 \includegraphics[width=\linewidth]{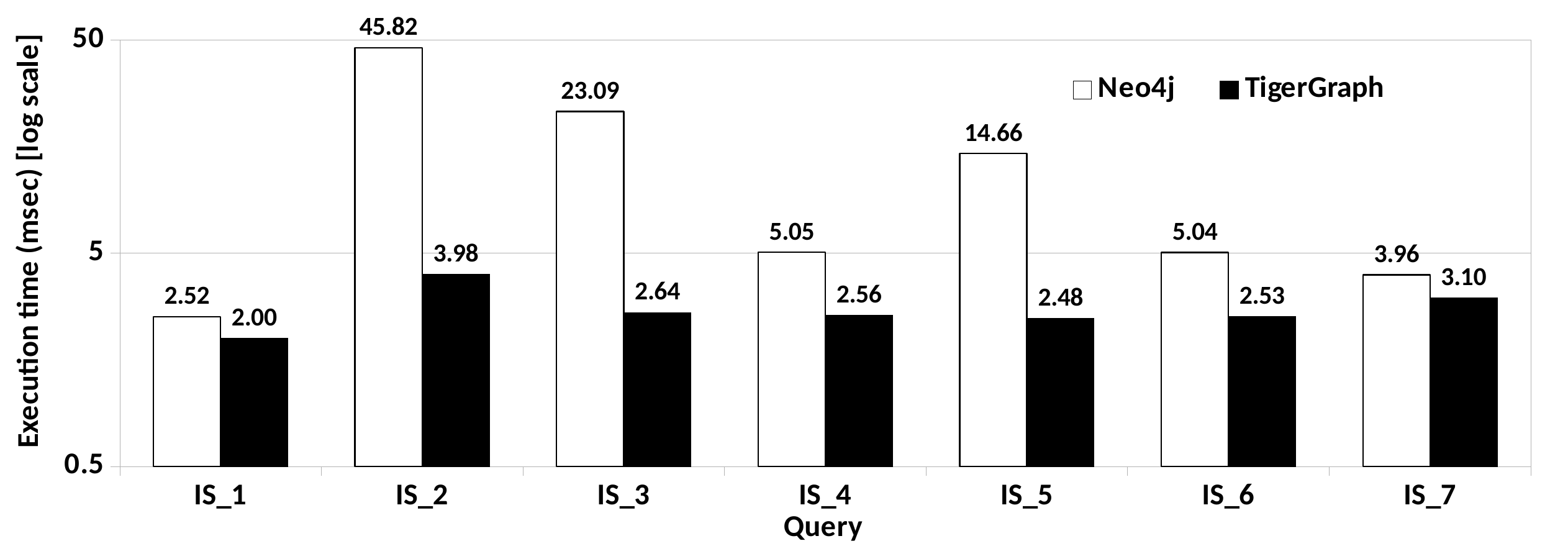}
 \caption{}
 \label{fig:sf-100-is}
\end{subfigure}

\begin{subfigure}[b]{.76\textwidth}
 \includegraphics[width=\linewidth]{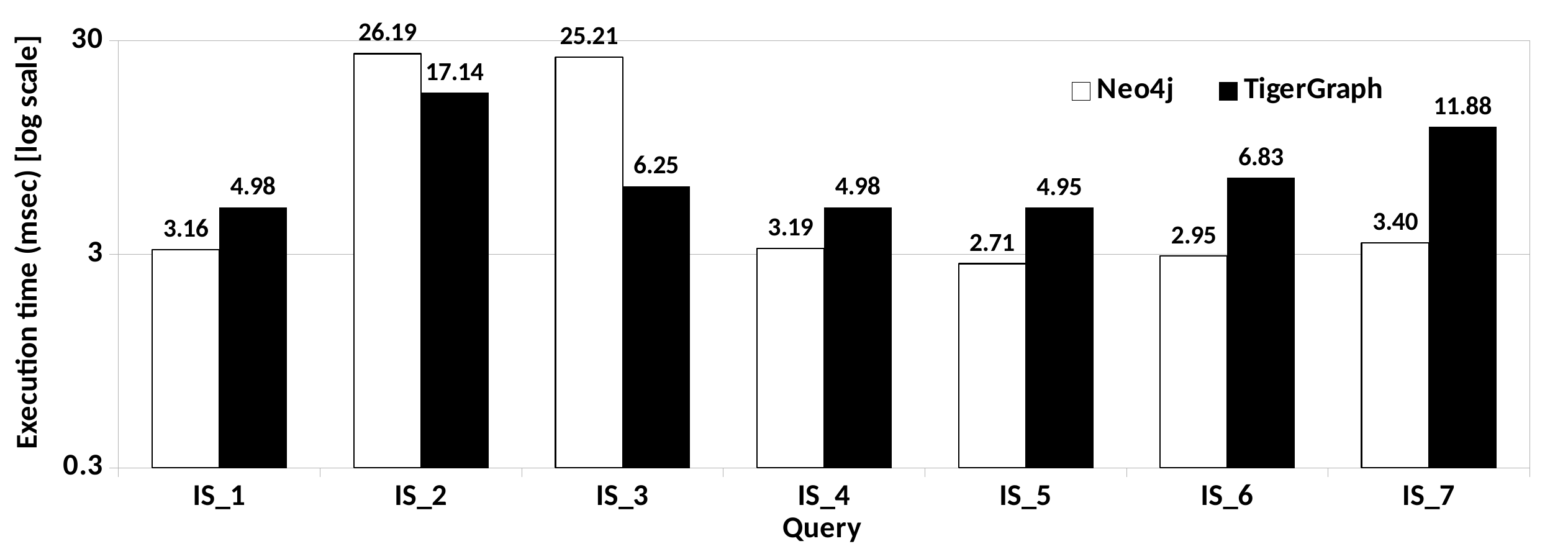}
 \caption{}
 \label{fig:sf-1000-is}
\end{subfigure}

\caption{Execution time in milliseconds (msec) for interactive short (IS) queries over scale factor 1 (a), 10 (b), 100 (c), and 1000 (d).}
\label{fig:sf-is}
\end{figure}
%%%%%%%%%%%%%%%%%%%%%%%%%

%%%%%%%%%%%%%%%%%%%%%%%%%%%%%%%%%%
\paragraph*{Query IS workload.}
Figure~\ref{fig:sf-is} depicts the runtime for the seven queries in the IS workload across all the four scale factors considered. Since all the queries access a very limited amount of data -- which is indexed on the search key -- the runtime is sub-second in all the cases. In fact, except queries IS\_2 and IS\_3, all the others run almost always in less than 10 milliseconds. This is in line with previously published results for other systems~\cite{snb-interactive,ozsu:rdbms-vs-graphs}. A careful reader immediately remarks that there is no clear relationship between the scale factor and the runtime---the runtime does not increase with the increase in the scale factor. Quite the opposite, there are queries for which the runtime decreases. This is because indexes are built and stored in memory for all the scale factors. As a result, the random memory accesses incur very similar time independent of the data size. Moreover, the graphs generated at larger scale factors are not supersets of smaller ones---an id that appears in SF-1 does not necessarily appears in SF-10. This forces us to modify the query parameters for every scale factor, which produces the observed variations. When comparing the two systems, TigerGraph is the clear winner for SF-1, SF-10, and SF-100---Neo4j is faster only for two queries. Then, somehow surprisingly, Neo4j outperforms TigerGraph for five queries at SF-1000. The only way we can explain this is by the properties of the machine on which the experiments are performed. Nonetheless, the difference between the two systems is rather inconsequential for this workload, given the small execution times. If this is the only workload someone is running, then any of the two systems is a good choice.

%%%%%%%%%%%%%%%%%%%%%%%%%
\begin{figure}[htbp]
\centering

\begin{subfigure}[b]{\textwidth}
 \includegraphics[width=\linewidth]{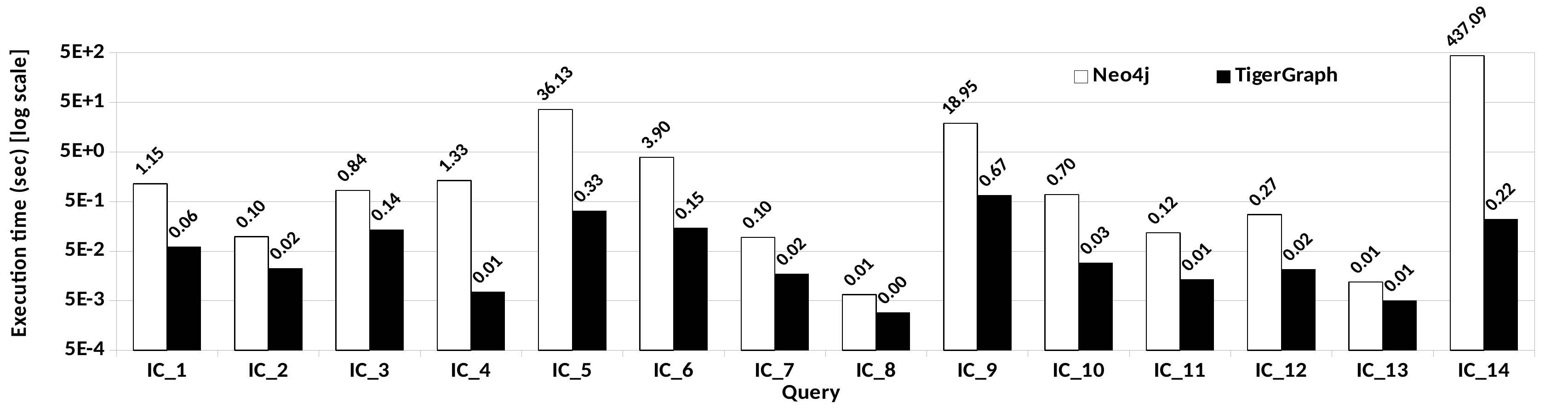}
 \caption{}
 \label{fig:sf-1-ic}
\end{subfigure}

\begin{subfigure}[b]{\textwidth}
 \includegraphics[width=\linewidth]{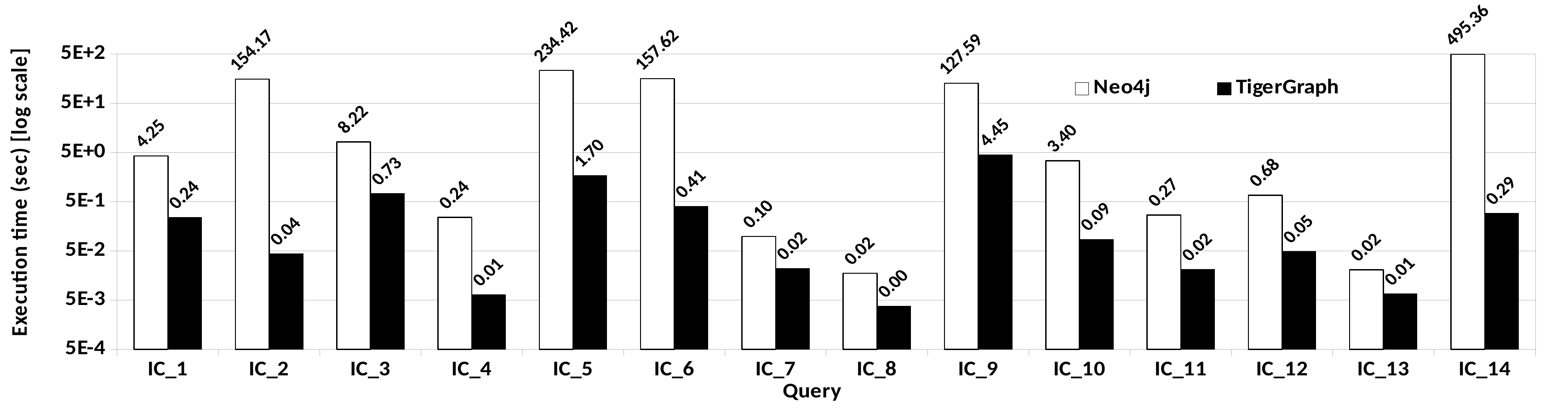}
 \caption{}
 \label{fig:sf-10-ic}
\end{subfigure}

\begin{subfigure}[b]{\textwidth}
 \includegraphics[width=\linewidth]{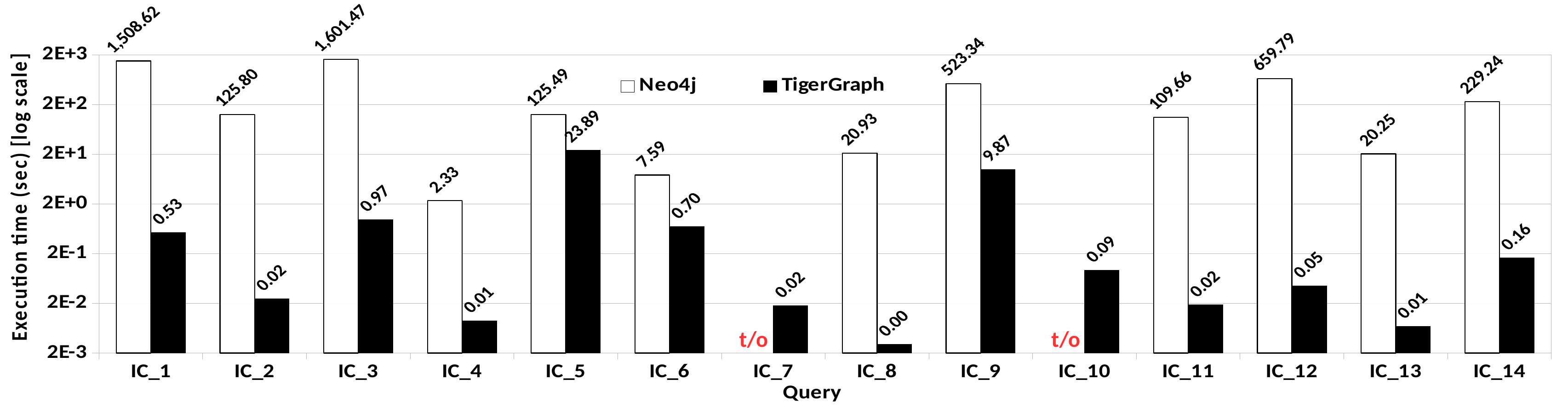}
 \caption{}
 \label{fig:sf-100-ic}
\end{subfigure}

\begin{subfigure}[b]{\textwidth}
 \includegraphics[width=\linewidth]{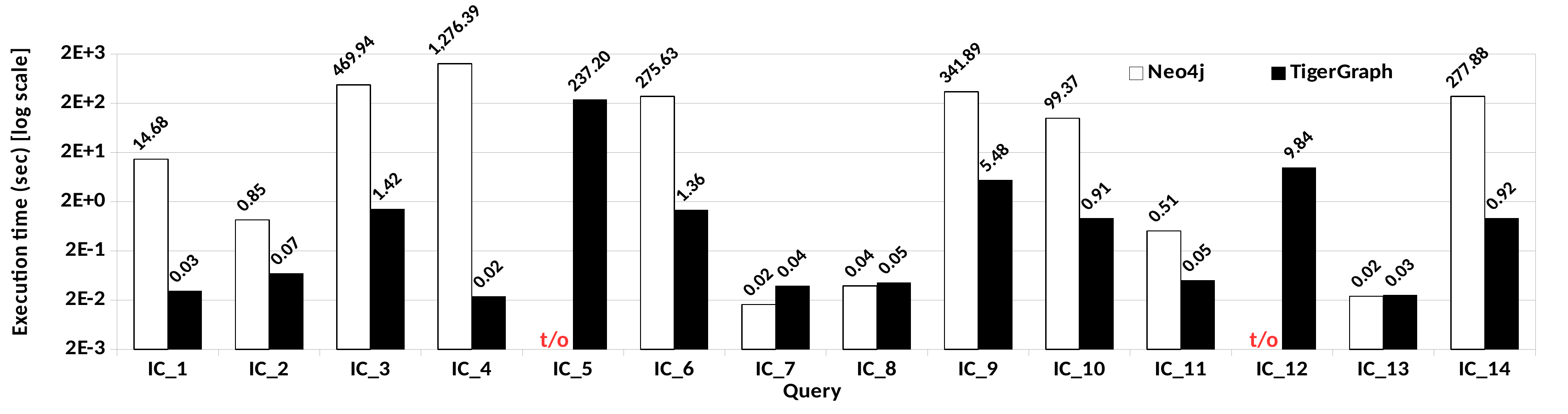}
 \caption{}
 \label{fig:sf-1000-ic}
\end{subfigure}

\caption{Execution time (sec) for interactive complex (IC) queries over scale factor 1 (a), 10 (b), 100 (c), and 1000 (d). t/o stands for timeout---execution did not finish in 18,000 seconds.}
\label{fig:sf-ic}
\end{figure}
%%%%%%%%%%%%%%%%%%%%%%%%%

%%%%%%%%%%%%%%%%%%%%%%%%%%%%%%%%%%
\paragraph*{Query IC workload.}
The situation is completely different for the IC workload depicted in Figure~\ref{fig:sf-ic}. There are only three queries across all the scale factors where Neo4j slightly outperforms TigerGraph. In all the other cases, TigerGraph is considerably faster---sometimes, by as much as four orders of magnitude. While TigerGraph finishes the queries in tens of milliseconds, they take more than a thousand seconds in Neo4j. Moreover, there are Neo4j queries that do not even finish execution at large scale factors in the allocated 5 hour timeout. Thus, TigerGraph is clearly the preferred choice for this workload. Since the amount of accessed data is larger and somewhat proportional with the scale factor, the runtime also generally increases with the scale factor. As for the indexing time, the increase is more accentuated for Neo4j than for Tigergraph. However, since the graphs are different, we have to change the query parameters and this sometimes results in a non-linear behavior.

%%%%%%%%%%%%%%%%%%%%%%%%%
\begin{figure}[htbp]
\centering

\begin{subfigure}[b]{\textwidth}
 \includegraphics[width=\linewidth]{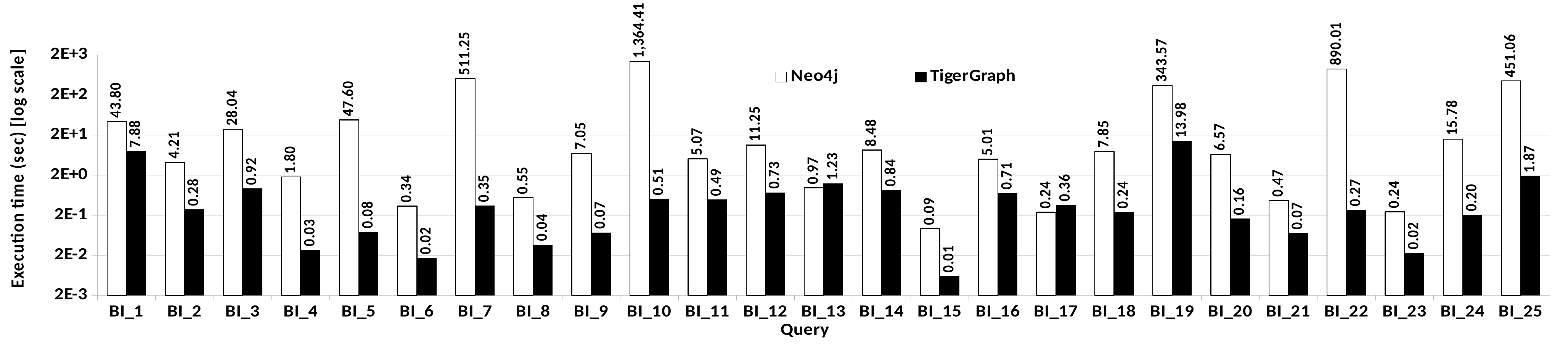}
 \caption{}
 \label{fig:sf-1-bi}
\end{subfigure}

\vspace*{1cm}

\begin{subfigure}[b]{\textwidth}
 \includegraphics[width=\linewidth]{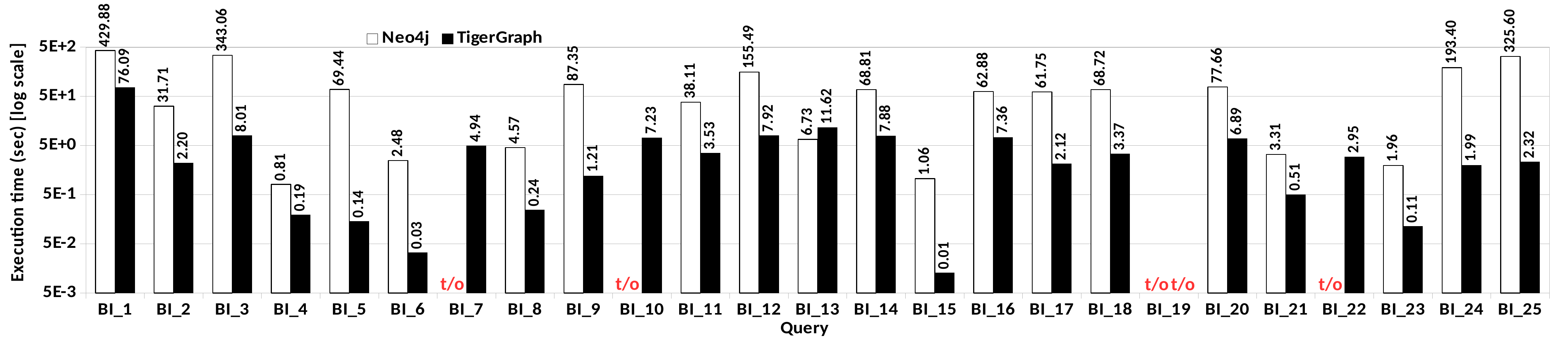}
 \caption{}
 \label{fig:sf-10-bi}
\end{subfigure}

\vspace*{1cm}

\begin{subfigure}[b]{\textwidth}
 \includegraphics[width=\linewidth]{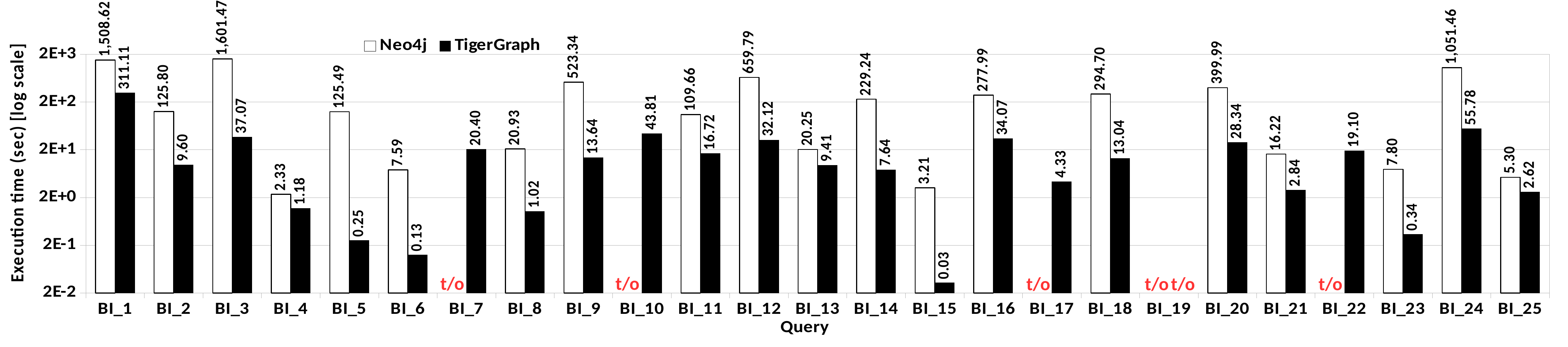}
 \caption{}
 \label{fig:sf-100-bi}
\end{subfigure}

\vspace*{1cm}

\begin{subfigure}[b]{\textwidth}
 \includegraphics[width=\linewidth]{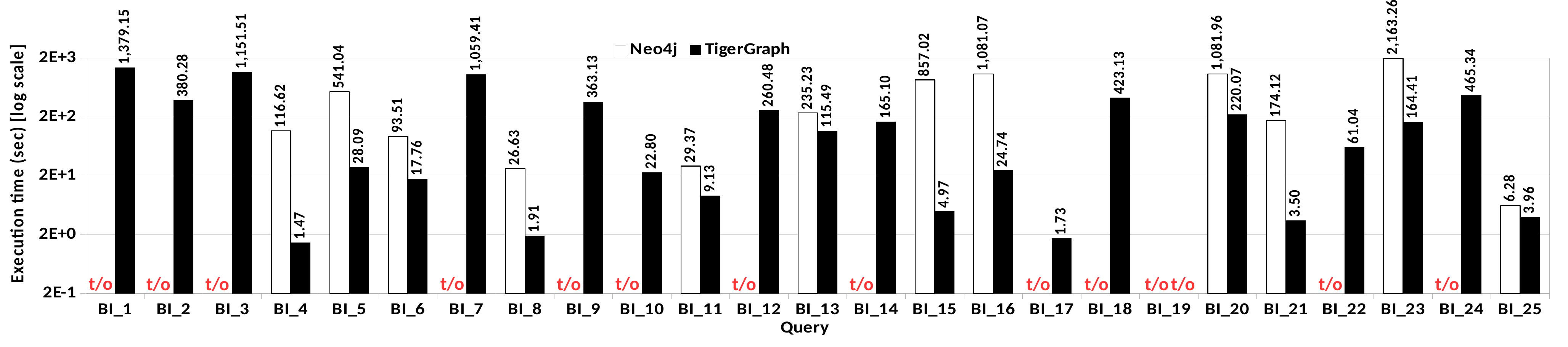}
 \caption{}
 \label{fig:sf-1000-bi}
\end{subfigure}

\caption{Execution time (sec) for business intelligence (BI) queries over scale factor 1 (a), 10 (b), 100 (c), and 1000 (d). t/o stands for timeout---execution did not finish in 18,000 seconds.}
\label{fig:sf-bi}
\end{figure}
%%%%%%%%%%%%%%%%%%%%%%%%%

%%%%%%%%%%%%%%%%%%%%%%%%%%%%%%%%%%
\paragraph*{Query BI workload.}
The results for the BI workload are shown in Figure~\ref{fig:sf-bi}. Again, TigerGraph clearly outperforms Neo4j across all the queries and all the scale factors---except two queries at SF-1 and one query at SF-10. On average, TigerGraph is close to one order of magnitude, i.e., 10X, faster than Neo4j across all the SF-100 and SF-1000 queries. Moreover, as the scale factor increases, fewer and fewer queries can be performed in the allocated time by Neo4j. For SF-1000, only 12 of the 25 queries finish execution before the timeout. This shows that Neo4j cannot scale to large data when performing complex BI queries. The results for TigerGraph are similar to the only other results published in the literature~\cite{snb-bi} up to scale SF-10. Of course, since this is not a direct comparison on the same hardware, it serves only as a basic guideline, not as an authoritative proof. We are not aware of any results for larger scale factors.

%%%%%%%%%%%%%%%%%%%%%%%%%%%%%%%%%%
\begin{table}[htbp]
  \begin{center}

  \resizebox{\textwidth}{!}{
		\begin{tabular}{l||rr|rr|rr|rr}

	\hline\hline
		
	\multirow{3}{*}{\textbf{Query}} & \multicolumn{8}{c}{\textbf{Execution time (seconds)}} \\
	\cline{2-9}
	& \multicolumn{2}{c|}{\textbf{SF-1}} & \multicolumn{2}{c|}{\textbf{SF-10}} & \multicolumn{2}{c|}{\textbf{SF-100}} & \multicolumn{2}{c}{\textbf{SF-1000}} \\
	& \textbf{TigerGraph} & \textbf{Neo4j} & \textbf{TigerGraph} & \textbf{Neo4j} & \textbf{TigerGraph} & \textbf{Neo4j} & \textbf{TigerGraph} & \textbf{Neo4j} \\
	\hline\hline

	IS\_1	& 0.0027	& \textbf{0.0023}	& 0.0026	& 0.1167	& 0.0020	& 0.0025	& 0.0050	& \textbf{0.0032} \\
	IS\_2	& 0.0143	& 0.1120	& 0.0159	& 0.0965	& 0.0040	& 0.0458	& 0.0171	& 0.0262 \\
	IS\_3	& 0.0214	& 0.0813	& 0.0174	& 0.0697	& 0.0026	& 0.0231	& 0.0062	& 0.0252 \\
	IS\_4	& 0.0032	& 0.0062	& 0.0026	& 0.0052	& 0.0026	& 0.0051	& 0.0050	& \textbf{0.0032} \\
	IS\_5	& 0.0030	& 0.0083	& 0.0029	& 0.0049	& 0.0025	& 0.0147	& 0.0049	& \textbf{0.0027} \\
	IS\_6	& 0.0034	& 0.0087	& 0.0031	& 0.0060	& 0.0025	& 0.0050	& 0.0068	& \textbf{0.0029} \\
	IS\_7	& 0.0080	& 0.0191	& 0.0062	& \textbf{0.0060}	& 0.0031	& 0.0040	& 0.0119	& \textbf{0.0034} \\

	\hline
								
	IC\_1	& 0.0613	& 1.1497	& 0.2412	& 4.2540	& 0.5317	& 1,508.6240	& 0.0307	& 14.6836 \\
	IC\_2	& 0.0226	& 0.0980	& 0.0439	& 154.1679	& 0.0249	& 125.7995	& 0.0696	& 0.8533 \\
	IC\_3	& 0.1357	& 0.8421	& 0.7330	& 8.2241	& 0.9703	& 1,601.4695	& 1.4237	& 469.9391 \\
	IC\_4	& 0.0075	& 1.3262	& 0.0065	& 0.2405	& 0.0090	& 2.3326	& 0.0237	& 1,276.3919 \\
	IC\_5	& 0.3257	& 36.1317	& 1.7019	& 234.4228	& 23.8868	& 125.4889	& 237.1967	& t/o \\
	IC\_6	& 0.1458	& 3.8971	& 0.4050	& 157.6176	& 0.6994	& 7.5908	& 1.3598	& 275.6304 \\
	IC\_7	& 0.0172	& 0.0950	& 0.0222	& 0.0987	& 0.0180	& t/o	& 0.0387	& \textbf{0.0163} \\
	IC\_8	& 0.0029	& 0.0066	& 0.0038	& 0.0176	& 0.0030	& 20.9287	& 0.0454	& \textbf{0.0391} \\
	IC\_9	& 0.6717	& 18.9488	& 4.4511	& 127.5934	& 9.8687	& 523.3391	& 5.4814	& 341.8940 \\
	IC\_10	& 0.0288	& 0.6950	& 0.0866	& 3.4005	& 0.0928	& t/o	& 0.9118	& 99.3705 \\
	IC\_11	& 0.0133	& 0.1167	& 0.0211	& 0.2693	& 0.0186	& 109.6632	& 0.0503	& 0.5141 \\
	IC\_12	& 0.0213	& 0.2718	& 0.0487	& 0.6813	& 0.0451	& 659.7865	& 9.8416	& t/o \\
	IC\_13	& 0.0051	& 0.0119	& 0.0067	& 0.0208	& 0.0069	& 20.2515	& 0.0251	& \textbf{0.0242} \\
	IC\_14	& 0.2188	& 437.0878	& 0.2927	& 495.3559	& 0.1640	& 229.2400	& 0.9184	& 277.8833 \\

		\hline
								
	BI\_1	& 7.8821	& 43.7982	& 76.0949	& 429.8811	& 311.1128	& 1,508.6240	& 1,379.1499	& t/o \\
	BI\_2	& 0.2766	& 4.2070	& 2.2038	& 31.7144	& 9.5975	& 125.7995	& 380.2766	& t/o \\
	BI\_3	& 0.9235	& 28.0367	& 8.0141	& 343.0650	& 37.0746	& 1,601.4695	& 1,151.5091	& t/o \\
	BI\_4	& 0.0270	& 1.7952	& 0.1929	& 0.8106	& 1.1765	& 2.3326	& 1.4743	& 116.6172 \\
	BI\_5	& 0.0755	& 47.5976	& 0.1433	& 69.4361	& 0.2512	& 125.4889	& 28.0933	& 541.0417 \\
	BI\_6	& 0.0170	& 0.3387	& 0.0330	& 2.4841	& 0.1263	& 7.5908	& 17.7641	& 93.5088 \\
	BI\_7	& 0.3453	& 511.2457	& 4.9366	& t/o	& 20.3964	& t/o	& 1,059.4062	& t/o \\
	BI\_8	& 0.0365	& 0.5524	& 0.2437	& 4.5672	& 1.0219	& 20.9287	& 1.9114	& 26.6252 \\
	BI\_9	& 0.0735	& 7.0492	& 1.2107	& 87.3451	& 13.6357	& 523.3391	& 363.1347	& t/o \\
	BI\_10	& 0.5125	& 1,364.4088	& 7.2261	& t/o	& 43.8146	& t/o	& 22.7974	& t/o \\
	BI\_11	& 0.4874	& 5.0742	& 3.5283	& 38.1115	& 16.7220	& 109.6632	& 9.1296	& 29.3734 \\
	BI\_12	& 0.7324	& 11.2471	& 7.9241	& 155.4858	& 32.1192	& 659.7865	& 260.4788	& t/o \\
	BI\_13	& 1.2253	& \textbf{0.9713}	& 11.6158	& \textbf{6.7296}	& 9.4079	& 20.2515	& 115.4907	& 235.2280 \\
	BI\_14	& 0.8399	& 8.4844	& 7.8764	& 68.8087	& 7.6439	& 229.2400	& 165.1050	& t/o \\
	BI\_15	& 0.0060	& 0.0927	& 0.0128	& 1.0641	& 0.0326	& 3.2056	& 4.9700	& 857.0237 \\
	BI\_16	& 0.7085	& 5.0081	& 7.3629	& 62.8762	& 34.0740	& 277.9866	& 24.7435	& 1,081.0724 \\
	BI\_17	& 0.3550	& \textbf{0.2371}	& 2.1243	& 61.7536	& 4.3261	& t/o	& 1.7270	& t/o \\
	BI\_18	& 0.2373	& 7.8512	& 3.3700	& 68.7250	& 13.0443	& 294.7010	& 423.1270	& t/o \\
	BI\_19	& 13.9819	& 343.5706	& t/o	& t/o	& t/o	& t/o	& t/o	& t/o \\
	BI\_20	& 0.1641	& 6.5713	& 6.8944	& 77.6586	& 28.3382	& 399.9949	& 220.0690	& 1,081.9585 \\
	BI\_21	& 0.0708	& 0.4676	& 0.5054	& 3.3095	& 2.8448	& 16.2169	& 3.5018	& 174.1151 \\
	BI\_22	& 0.2688	& 890.0080	& 2.9528	& t/o	& 19.0996	& t/o	& 61.0383	& t/o \\
	BI\_23	& 0.0226	& 0.2429	& 0.1133	& 1.9604	& 0.3400	& 7.8020	& 164.4104	& 2,163.2589 \\
	BI\_24	& 0.2008	& 15.7829	& 1.9883	& 193.4047	& 55.7750	& 1,051.4582	& 465.3403	& t/o \\
	BI\_25	& 1.8657	& 451.0648	& 2.3218	& 325.5987	& 2.6213	& 5.2952	& 3.9553	& 6.2800 \\

	\hline\hline
    \end{tabular}
  }
  
  \end{center}
\caption{Execution time (in seconds) for all the queries and all the scale factor data. t/o stands for timeout---execution did not finish in 18,000 seconds. The bold values correspond to the cases when Neo4j is faster than TigerGraph.}\label{tbl:exec-time-all}
\end{table}
%%%%%%%%%%%%%%%%%%%%%%%%%%%%%%%%%%

%%%%%%%%%%%%%%%%%%%%%%%%%%%%%%%%%%
\subsection{Summary}\label{sec:experiments:summary}

We can summarize the results of our in-depth experimental study as follows:
\begin{compactitem}
\item TigerGraph stores graph data considerably more compactly than Neo4j. It uses 3X less storage for the raw data and 4X less storage if we include the indexes. Moreover, TigerGraph compresses the original raw data generated by the LDBC SNB data generator by a factor of 2X.
\item Neo4j is faster at ingesting raw data than TigerGraph. The difference is 3X for SF-1 and decreases with the increase in scale factor. However, Neo4j has a rather non-scalable index building algorithm which grows the indexing time at an exponential rate. As a result, for SF-1000, TigerGraph achieves a total loading time that is 2X faster than in Neo4j.
\item Table~\ref{tbl:exec-time-all} summarizes the runtime for all the 46 queries in the SNB workloads across all the four scale factors. Out of the 368 configurations, Neo4j is faster than TigerGraph only in 13 cases---bold in the table. This represents 3.5\% of the workload. Thus, it is clear that TigerGraph is superior to Neo4j on the LDBC SNB benchmark.
\end{compactitem}

%%%%%%%%%%%%%%%%%%%%%%%%%%%%%%%%%%
\section{CONCLUSIONS AND FUTURE WORK}\label{sec:conclusions}

In this study, we present the first results of a complete implementation of the LDBC SNB benchmark in two native graph database systems---Neo4j and TigerGraph. In addition to thoroughly evaluating the performance of all of the 46 queries in the benchmark on four scale factors and three computing architectures, we also measure the bulk loading time and storage size. Our results show that TigerGraph is consistently outperforming Neo4j on the vast majority of the queries---more than 95\% of the workload. The gap between the two systems increases with the size of the data since only TigerGraph is able to scale to SF-1000---Neo4j finishes only 12 of the 25 BI queries in reasonable time. Nonetheless, Neo4j is generally faster at bulk loading graph data -- if we ignore the index building time -- and has a more compact declarative query language. In order to encourage reproducibility, we make all the code, scripts, and configuration parameters publicly available online~\cite{zhiyi-code,tigergraph-ldbc-snb-queries}. In the future, we plan to include more systems in the study, both graph and relational databases.

%%%%%%%%%%%%%%%%%%%%%%%%%
\paragraph*{Acknowledgments.}
We would like to thank TigerGraph, Inc. for the support they have provided for this work. This support has come in two forms. First, TigerGraph engineers have been actively involved in the optimization of the SNB workloads in GSQL. Second, Zhiyi Huang has been financially supported by TigerGraph funds. Nonetheless, the findings presented in this report are the sole contribution of the authors.

%%%%%%%%%%%%%%%%%%%%%%%%%%%%%%%%%%%%%%%%%%%%%%%%%%%%%%%
\bibliographystyle{abbrv}
%\bibliography{biblio}

%%%%%%%%%%%%%%%%%%%%%%%%%%%%%%%%%%%%%%%%%%%%%%%%%%%%%%%

\end{document}